\documentclass[12pt]{article}
		\title{Estimating Design Operating Characteristics in Bayesian Adaptive Clinical Trials}

\pdfoutput=1
\date{}
	\usepackage{amsmath,latexsym, graphics,graphicx,showexpl,amsthm,amsfonts,fullpage}
\usepackage[left=1in,top=1in,right=1in,nohead]{geometry}
\usepackage{graphicx}
\usepackage[margin=10pt,font={small},labelfont=bf]{caption}
\usepackage{subcaption}
\usepackage{bbm} 
\usepackage{setspace}
\usepackage{natbib}
\usepackage[titletoc]{appendix}
\usepackage{epstopdf}
\usepackage[pdftex,bookmarks,colorlinks=true,linkcolor=blue]{hyperref}
\usepackage{algpseudocode,algorithm}
\usepackage{authblk}
\usepackage{multicol}
\usepackage{hyperref}
\hypersetup{colorlinks,
	citecolor=blue
}
\usepackage{array}
\usepackage{etoolbox}
\usepackage{setspace}
\usepackage{rotating}

\newcolumntype{L}[1]{>{\raggedright\let\newline\\\arraybackslash\hspace{0pt}}m{#1}}
\newcolumntype{C}[1]{>{\centering\let\newline\\\arraybackslash\hspace{0pt}}m{#1}}
\newcolumntype{R}[1]{>{\raggedleft\let\newline\\\arraybackslash\hspace{0pt}}m{#1}}
\AtBeginEnvironment{thebibliography}{\linespread{1.5}\selectfont}

\makeatother
\author[1]{Shirin Golchi\thanks{shirin.golchi@mcgill.ca}}
\affil[1]{Department of Epidemiology, Biostatistics and Occupational Health, McGill University, Montreal, QC H3A 1G1, Canada}

\begin{document}

	\maketitle

\begin{abstract}
	Bayesian adaptive designs have gained popularity in all phases of clinical trials with numerous new developments in the past few decades. During the COVID-19 pandemic, the need to establish evidence for the effectiveness of vaccines, therapeutic treatments and policies that could resolve or control the crisis emphasized the advantages offered by efficient and flexible clinical trial designs. In many COVID-19 clinical trials, due to the high level of uncertainty, Bayesian adaptive designs were considered advantageous. Designing Bayesian adaptive trials, however, requires extensive simulation studies that are generally considered challenging, particularly in time-sensitive settings such as a pandemic. In this article, we propose a set of methods for efficient estimation and uncertainty quantification for design operating characteristics of Bayesian adaptive trials. Specifically, we model the sampling distribution of Bayesian probability statements that are commonly used as the basis of decision making. To showcase the implementation and performance of the proposed approach, we use a clinical trial design with an ordinal disease-progression scale endpoint that was popular among COVID-19 trial. However, the proposed methodology may be applied generally in clinical trial context where design operating characteristics cannot be obtained analytically.
\end{abstract}

\noindent%

\section{Introduction}
Adaptive designs are one category of flexible alternatives to conventional fixed-size randomized clinical trials (RCTs). In adaptive designs,  decisions to stop or adapt may be made at interim analyses according to accumulating evidence which, in some cases, can result in reduced sample sizes and cost. In addition, participants benefit from an increased chance of receiving a beneficial treatment. While adaptive designs in clinical trials go beyond Bayesian adaptive trials  \citep{PalBedCho18, BurMoxPal20}, Bayesian methods in adaptive trials have become popular. The main reasons for this popularity are that the Bayesian framework naturally accommodates sequential analysis of accumulating data and that the validity of Bayesian probability statements is not affected by small sample sizes or incorporation of  stopping rules \citep{BerryBook}.

There have been major developments in the design of Bayesian adaptive trials during the past few decades \citep{Ber89, BerEic95, CarKadGel98}. Specifically these designs have been widely applied to drug development \citep{ThaCooEst06, MueBerGri06} and disease specific fields such as cancer trials \citep{BuzIbrFra05, ZhoLiuKim08, ISPY2, BisLiuLee09}.

Despite their growing use of Bayesian methods in clinical trials, proposed Bayesian adaptive designs are primarily assessed by regulatory agencies based on their frequentist operating characteristics such as power and false positive rate.  The Food and Drug Administration (FDA), for example, emphasizes the importance of simulation studies in evaluating operating characteristics in adaptive trials of drugs and biologics \citep{FDA_ DOC}.

Various stopping rules may be incorporated  into a Bayesian adaptive trial design with multiple interim analyses. Common decision rules include stopping the trial for efficacy or futility, or eliminating an arm if the probability of effectiveness is small for the corresponding treatment. Stopping rules are most commonly defined based on a Bayesian test statistic derived from the posterior distribution of model parameters. These include the posterior or posterior predictive probability of effectiveness. i.e. the alternative hypothesis. Efficacy or futility are then defined based on  high and low thresholds for these Bayesian probability statements.  Another common adjustment is response adaptive randomization, where allocation ratios are adjusted with respect to posterior or posterior predictive probability statements.

Similar to the critical region of a frequentist test statistic, thresholds for Bayesian probabilities can be specified to achieve satisfactory design operating characteristics ( DOCs). Power, for instance, may be defined as the probability of observing a high posterior probability of effectiveness given an assumed effect size. The sampling distribution of the posterior or posterior predictive probability statement (Bayesian test statistic) is therefore required to obtain power. Unlike in the classical hypothesis testing framework, however, the sampling distribution of a Bayesian probability statement is not available in closed form. Therefore,  DOCs are typically estimated via Monte Carlo methods, i.e., by simulating the trial and the sampling distribution of the Bayesian test statistic for a given set of parameters.

Simulation studies for design of Bayesian adaptive trials can be time consuming since the combination of plausible ranges of model parameters, includingeffect size and baseline measure assumptions, together with possible design parameters, including efficacy and futility thresholds, sample sizes and frequency of interim analyses, can result in a large number of simulation scenarios. In many cases, a complex trial design and/or analysis model without analytically tractable posteriors requires a significant amount of computation --- involving methods such as Markov chain Monte Carlo (MCMC) approaches --- for a single trial simulation which is multiplied by the number of simulation scenarios as well as by the number of iterations. 

The methods proposed in the present article aremotivated by a clinical trial design exercise in the context of the COVID-19 pandemic. A brief introduction to this clinical trial is provided in Section~\ref{Sec:trial}. In early discussions among the team of  investigators, the ordinal scale disease progression endpoint, as defined by the World Health Organization (WHO) \citep{WHO20}, was considered as the primary endpoint of the trial with the goal of evaluating the effect of the intervention in reducing the severity of COVID-19. The Proportional Odds (PO) ordinal logistic regression was considered as the analysis model which is one of the simplest statistical models used for the analysis of the ordinal scale outcome \citep{HarLin20}. Bayesian inference for the PO model, however, requires MCMC sampling since the analytic form of the posterior is not available. Therefore, assessing DOCs for Bayesian trials with the ordinal scale outcome requires extensive simulations and carries a significant computational burden. The computational requirements, together with the uncertainty of the risk associated with the levels of the ordinal scale outcome, resulted in a simplification of the design that used a binary endpoint. This shifted the focus to the risk of infection rather than disease severitywhich was a necessary change to meet deadlines for funding applications and regulatory approvals dictated by the time-sensitive situation. 

As a motivating example, we consider the hypothetical case with the ordinal scale disease progression outcome and the PO model used in the design of the above COVID-19 trial. The link between COVID-19 infection and  disease severity may have been strong prior to vaccination. In the absence of a strong association, however, changing the primary outcome for computational reasons would not have been acceptable and could have impacted the patient outcomes and the efficiency or appeal of the study. The present work will facilitate choosing the clinically correct endpoint by addressing computational feasibility.

In this article, we propose a set of methods to overcome the computational hurdles in evaluating DOCs as well as assessing the sensitivity of  DOCs to the model and design parameters in Bayesian adaptive trials. The methodology can be employed in general for design of clinical trials where power analysis and assessing DOCs relies on simulations rather than analytic forms for the sampling distribution of the test statistic. The proposed approach is to estimate the sampling distribution of Bayesian probability statements i.e., the test statistics) through the model parameter space which is then used to provide computationally efficient estimates of  DOCs for a wide range of assumptions and decision parameters without the need for additional simulations. A simple parametric density estimation approach is employed where the sampling distribution of the probability statement is assumed to follow a Beta distribution whose parameters are modelled as Gaussian processes (GP) with distance-based correlation structure across the parameter space. The GPs are trained over an initial set of simulated distributions for the test statistic at a select set of parameter values.

The novelty of the present article is in proposing a set of methodology for the design of Bayesian adaptive trials beyond simple Monte Carlo simulations for the assessment of  DOCs. Currently, methods for estimating the operating characteristics for a given set of parameters rely solely on simulations and are not accompanied with adequate uncertainty quantification. The present work sets the foundation for development of methods that facilitate the use of Bayesian measures for decision making in clinical trials while satisfying traditional requirements. 

As mentioned above, the ordinal scale outcome together with the PO model is used to showcase the implementation and performance of the proposed approach. One interesting but challenging aspect of using the ordinal scale outcome is incorporating the uncertainty associated with the risks for disease severity levels in the trial design. This is addressed by exploring the  DOCs over a range of plausible risk vectors. The vector of probabilities must add up to one which results in a simplex. Moreover, in GP-based models, satisfactory prediction performance throughout the input (parameter) space, is achieved by spreading the initial set of target function evaluations uniformly using a space-filling design \citep{OHa78, McKBecCon79, SacWelMit89, JohMooYlv1990, MorMit95}. Space-filling designs over non-rectangular, constrained spaces have recently received considerable attention \citep{LinShaWin10, LekJon2014, MakJos18}.  Generating a space--filling design on a simplex, however, is not straightforward. We use the method of \cite{LekJon2014}, which generates a covering sample of the target space, clusters the sample points and selects the cluster centroids as the design points. We propose a sampling technique for generating an initial sample on a simplex as an efficient and effective alternative to simple Monte Carlo.

The remainder of this article is organized as follows. The clinical trial design that motivates the development of the proposed approach is described in Section~\ref{Sec:trial}. In Section~\ref{Sec:methods}, we introduce a method for emulating and predicting the sampling distribution of a Bayesian test statistic using a beta prior distribution with GP parameters. We apply the proposed approach to the motivating example in Section~\ref{Sec:PO}. We also discuss the design of the training set for this application and assessment of estimation using cross-validation. Section~\ref{Sec:dis} follows with a discussion of the limitations of the proposed approach and future extensions.

\section{Motivating Example}
\label{Sec:trial}
This work was motivated by the design of a clinical trial investigating the effectiveness of high-dose vitamin D in preventing severe cases of COVID-19 (PRevention of COVID-19 with high dose Oral Vitamin D supplemental Therapy in Essential healthCare Teams---PROTECT). Given the goal of the study, the ordinal scale disease progression outcome recommended by the WHO \citep{WHO20} was of interest. The clinical progression scale has ten disease progression level from uninfected to dead. The challenge in use of a ten-level endpoint in clinical trials is the lack of granular data to inform the base rate assumptions as well as the dimensionality of the parameter space in statistical inference. Therefore, simplifications of the 10-level ordinal scale that merge levels may be used. Table~\ref{tab:OS} shows the definition of the five-level ordinal scale disease progression endpoint with categories uninfected, mild disease, moderate disease, severe disease and dead. 

\begin{table}
	\caption{Levels and description of the ordinal scale disease-severity endpoint.}
	\begin{tabular} {l | p{8.5cm} | c}
		Patient State & Descriptor & Level \\
		\hline\hline
		Uninfected & Uninfected; no viral RNA detected & 0\\
		\hline
		Ambulatory mild disease & Asymptomatic; viral RNA detected  & 1\\
		& Symptomatic; independent &\\
		& Symptomatic; assistance needed& \\
		\hline
		Hospitalised: moderate disease & Hospitalised; no oxygen therapy & 2\\
		& Hospitalised; oxygen by mask or nasal prongs & \\
		\hline
		Hospitalised: severe disease & Hospitalised: oxygen by NIV or high flow & 3\\
		&Intubation and mechanical ventilation &\\
		\hline
		Dead & Dead & 4\\
		
	\end{tabular}
	\label{tab:OS}
\end{table}

While the ordinal scale endpoint was considered in early stages of the design of PROTECT, due to the scale of simulation studies required to assess the DOCs and the level of uncertainty in the base risks associated with each outcome level, the focus was changed to the risk of infection and a binary primary outcome (incidence of laboratory-confirmed COVID-19 infection) that would accommodate closed-form posterior updates. The PROTECT study adopted a Bayesian adaptive design that allowed for an interim analysis when 75\% of (approximately 1000) patients completed their follow-up time (16 weeks). The total sample size was estimated via simulations to achieve at least 80\% probability of success. The trial could be concluded for efficacy at the interim analysis in case of sufficient evidence in favour of effectiveness. The control risk of infection was also monitored and the data up to the interim analysis would be used to obtain the posterior predictive power at the final analysis. If the posterior predictive power was below the 80\% target, the study would be prolonged for up to 24 weeks of follow-up in order to achieve a higher chance of success. The interim analysis and the adaptive components were essential due to the quickly evolving situation of the pandemic. The PROTECT study was funded by the CIHR COVID-19 Rapid Research Funding Opportunity approved by Health Canada and started recruiting in February 2021 (ClinicalTrials.gov Identifier: NCT04483635). However, the launch of the study coincided with the widespread distribution of vaccines to health care workers, which resulted in a significant drop in recruitment rates and, eventually, termination of the trial due to lack of feasibility. 

In Section~\ref{Sec:PO}, we consider the hypothetical case where the ordinal scale outcome is used as the primary outcome in the PROTECT design and argue that the proposed techniques in this article can improve the design of adaptive clinical trials by enabling the use of outcomes of interest with more realistic statistical models while accounting for the uncertainty associated with parameter assumptions in a computationally efficient framework.

We focus on the probability of stopping early for superiority at the interim analysis with 75\% of participant outcomes as the  DOCs of interest in order to illustrate the implementation and performance of the proposed approach. We emphasize that implementation and performance do not necessarily depend on the specific trial design or the specific  DOCs of interest.

\section{Methodology}
\label{Sec:methods}
Statistical success or significance in Bayesian adaptive clinical trials is commonly defined based on posterior or posterior predictive probability of the alternative hypothesis \citep{BerryBook}. Stopping decisions are made according to the same criterion. For example, the decision of stopping the trial for efficacy at an interim analysis may be made if the posterior probability of effectiveness exceeds a prespecified upper probability threshold $U$ (e.g., $U = 0.95$), that is, if
\begin{equation}
	\label{ps}
	\pi(\mathbf{y}_t) = P(H_A\mid \mathbf{y}_t)>U,
\end{equation}
where $H_A$ is the alternative hypothesis that is often formulated as the treatment having at least a certain magnitude of effect and $\mathbf{y}_t$ denotes the participants' outcomes accumulated up to the decision time $t$. Therefore, $\pi(\mathbf{y}_t)$ can be viewed as a test statistic and $U$ is the critical value with respect to which statistical significance is determined. We use $\pi$ as general notation for a Bayesian test statistic used in a generic Bayesian adaptive trial.

% Effectiveness is of course defined within the context. For example, in the PROTECT trial, the ratio of the odds of severe disease between the treatment and control arm was the estimand and the alternative hypothesis is $H_A: OR<1$. Therefore, the corresponding test statistic is $P(OR<1\mid \mathbf{y})$.

Within the conventional RCT framework, the known sampling distribution of the test statistic under the null and alternative hypotheses are used to specify critical values that achieve target  DOCs such as a 5\% false-positive rate and 80\% power. The sampling distribution of a posterior probability statement $\pi$, however, is not available in general. Therefore, DOCs are typically estimated using Monte Carlo simulations for a given set of model parameters that correspond to points under the null and alternative hypotheses as well as a range of critical values. 

Evaluating $\pi$ for every simulated set of $\mathbf{y}$ may require MCMC sampling except for simple models where a conjugate prior is available. While this is not generally a hurdle in Bayesian inference given the vast number of sampling algorithms, approximation methods and computational advancements in the literature, involved Bayesian computation can within a simulation study can be impractical. For example, estimating  DOCs in a trial design with $I$  (interim and final) analyses, at $n_T$ sets of model parameter values and $L$ decision thresholds via  $M$ simulation iterations and $J$ MCMC iterations per evaluation of $\pi$ has a computational complexity of $\mathcal{O}(n_T  I M J  L)$. In the following we propose an approach that reduces this computational complexity to $\mathcal{O}(n_t I M J  )$ where $n_t \ll n_T$ and the range of decision threshold values does not contribute to the computational complexity.

Denote the model parameter space,  including the plausible range of all baseline and treatment effect parameters, by $\Theta$. The null and alternative hypotheses define a partition of the parameter space, $\Theta = \Theta_0 \cup \Theta_A$ ( where $H_0: \theta \in \Theta_0$ and $H_A:\theta \in \Theta_A$). In simulation studies designed to asses operating characteristics of a clinical trial design, a variety of parameter values within $\Theta$ that correspond to various parameter points under the null and alternative hypotheses need to be explored. In addition to power and false positive rate, a variety of other  DOCs are of interest in adaptive designs. For example, power at a given interim analysis, i.e., the probability of stopping the trial early due to a correct efficacy conclusion, is 
\begin{equation}
	\label{eq:ip}
	P_y(\pi_\text{int}>U \mid \theta^* \in \Theta_A),
\end{equation}	
where $\pi_\text{int}$ is the Bayesian test statistic at the interim point and the subscript $y$ denotes the probability under the data or sampling distribution. Therefore, the sampling distribution of $\pi_\text{int}$ over the parameter space $\Theta$ is key to assessing  DOCs. 
 
We propose a model for the sampling distribution of $\pi$ that allows us to ``learn" the distribution function of $\pi$ over $\Theta$ through the empirical sampling distribution obtained at a small set of parameter values. We assign the following prior distribution to $\pi$ at a given point $\theta \in \Theta$:
\begin{equation}
\label{eq:beta}
\pi\mid \theta \sim \text{Beta}\left(a(\theta), b(\theta)\right)
\end{equation}
with
\begin{equation}
\label{GPprior}
a(\theta) \sim \mathcal{GP} (\mu_a, \rho_a(\theta, \theta')) \hskip 10pt \text{and} \hskip 10pt
b(\theta) \sim \mathcal{GP} (\mu_b, \rho_b(\theta, \theta')),
\end{equation}
where $\mathcal{GP} (\mu, \rho(\theta, \theta'))$ denotes a GP with a constant mean $\mu$ and a covariance function $\rho(\theta, \theta')$. The mean and covariance parameters of the GP prior distribution are trained independently for the shape and scale parameters of the beta distribution in (\ref{eq:beta}). 

The GP priors in (\ref{GPprior}) are based on the assumption that the parameters of the target distribution $f(\pi\mid \theta)$ are smooth in $\Theta$. This, in turn, results in predictions for DOCs that are also smooth in $\Theta$. We take the covariance functions of the GP processes in (\ref{GPprior}) to be the squared exponential covariance function, which is one of the most common choices in GP modelling \citep{RasWil2006}. This choice of covariance function assumes infinite differentiability with respect to $\theta$. This may prove an unrealistic assumption in some applications and may give rise to convergence issues when estimating GP parameters. However, we do not encounter such issues in the application described in Section~\ref{Sec:PO}. 

The GPs in~(\ref{GPprior}) are trained over the parameter space using estimates of $a(\theta)$ and $b(\theta)$ at select $\theta\in\Theta$. Let $\boldsymbol{\theta}_t$ denote the training set of size $n_t$. The trial is simulated for $M$ iterations according to the specified design for the parameter values in $\boldsymbol{\theta}_t$, which results in a Monte Carlo sample of the distribution of $\pi$. A beta distribution is fit over the Monte Carlo sample at each training point using the method of moments to obtain estimates of $a(\theta)$ and $b(\theta)$ that match the mean and variance of the empirical distribution. These estimates, $\hat{\mathbf{a}} = \hat{a}(\boldsymbol{\theta}_t)$ and $\hat{\mathbf{b}} = \hat{b}(\boldsymbol{\theta}_t)$ are then used as realizations to obtain the posterior GPs for any parameter values throughout $\Theta$:

\begin{equation}
	\label{GPposterior}
	a(\theta^*)\mid \hat{\mathbf{a}} \sim \mathcal{N} (\mu^\text{post}_a, V(a(\theta^*))) \hskip 10pt \text{and} \hskip 10pt
	b(\theta^*)\mid \hat{\mathbf{b}} \sim \mathcal{N} (\mu^\text{post}_b, V(b(\theta^*))),
\end{equation}

where
\begin{equation*}
\mu^\text{post}_a = \mu_a + \mathbf{k}_a^\top (K_a+\sigma_a^2I)^{-1}\mathbf{a}, \hskip 10pt
\mu^\text{post}_b = \mu_b + \mathbf{k}_b^\top(K_a+\sigma_b^2I)^{-1}\mathbf{b},
\end{equation*}
\begin{equation*}
	V(a(\theta^*)) = \mathbf{k}_a(\theta^*, \theta^*) + \mathbf{k}_a^\top(K_a+\sigma_a^2I)^{-1}\mathbf{k}_a
\end{equation*}
 and 	\begin{equation*} 
V(b(\theta^*)) = \mathbf{k}_b(\theta^*, \theta^*) + \mathbf{k}_b^\top(K_b+\sigma_b^2I)^{-1}\mathbf{k}_b.
\end{equation*}
The $\mathbf{k}$s are vectors of size $n_t$ whose components are  $\rho(\theta_t, \theta^*)$, the $K$s are covariance matrices whose components are $\rho(\theta_t, \theta_{t'})$, and the $\sigma^2$s are the observation variances estimated for each GP model.

Once the posterior GPs are obtained, the sampling distribution of $\pi$ is fully specified (predicted) at any point in $\Theta$ and any  DOCs of interest may be evaluated as a tail probability of the beta distribution. For example, the interim power in~(\ref{eq:ip}) is the 100$U$\% upper tail probability of a Beta distribution whose parameters are given as in~(\ref{GPposterior}). The posterior uncertainty of these parameters will translate into uncertainty estimates for the corresponding tail probability, i.e., the  DOCs estimates.

The specification of the training set $\boldsymbol{\theta}_t$ is important in the performance of the proposed model, as in any predictive-modelling framework. There exists much literature on GP modelling and the design of computer experiments regarding techniques for constructing training sets on a parameter (input) space. Specifically, space-filling designs are recommended to optimize predictive performance \citep{OHa78, McKBecCon79, SacWelMit89, JohMooYlv1990, Tan93, MorMit95, Ye98, SanWilNot03, JinCheSud05, JosGul15}. For a comprehensive review, see \cite{Jos16} and the references therein. In the next section, we apply the proposed approach to a specific  DOC estimation example, discuss the construction of a space-filling design on a nonrectangular, nonconvex parameter space and propose a simple design algorithm.

\section{Application to the PROTECT Design}
\label{Sec:PO}
In this section, we consider the design of the PROTECT study and a hypothetical scenario that the ordinal scale disease progression endpoint is the primary outcome. The primary analysis is performed using a PO model inspired in \cite{HarLin20}. This reference also provides a  comprehensive discussion of the design of Bayesian adaptive trials with ordinal-scale outcomes. 

For the  DOC of interest, we will focus on the probability of stopping early for efficacy or futility at the interim analysis with 75\% of participant outcomes. Without loss of generality, we consider a four-level outcome. We apply the model proposed in the previous section to estimate the data distribution of the posterior probability of effectiveness throughout the model parameter subspace that corresponds to a set of credible base risks and effect assumptions under the null and alternative hypotheses. Then we estimate the probability of stopping early for a number of decision thresholds.

An individual outcome is assumed to follow a multinomial distribution,
$Y \sim \text{Multinom}(1, \mathbf{p} = (p_1, \ldots, p_4))$, 
where $p_k$ ($k = 1, \ldots, 4$) is the risk associated with the $k^\text{th}$ level of disease severity and $\sum_{k = 1}^4 p_k = 1$. The PO model is a logistic regression of the tail probability,
\begin{equation}
	\label{eq:logistic}
	P(Y\geq k \mid A) = \frac{1}{1+\exp[-(\alpha_k + \beta A)]} \hskip 30pt \text{for}  \hskip 5pt k = 2, 3, 4.
\end{equation}
Note that $P(Y\geq 1 \mid A)=1$ and
\begin{equation*}
	\alpha_k = -\log \frac{\sum_{i = 1}^{k-1}p_i}{1-\sum_{i = 1}^{k-1}p_i}.
\end{equation*}
In Equation~(\ref{eq:logistic}), $A$ is the treatment indicator, i.e., $A = 1$ indicates that the patient has received the treatment and $A=0$ indicates assignment to the control arm. The ratio of the odds of disease severity is then represents the effect of the treatment:
\begin{equation*}
	\mathrm{OR} = \frac{P(Y\geq k\mid A = 1)}{P(Y\geq k\mid A = 0)} = \exp(\beta).
\end{equation*}
This simple parametrization reduces the treatment effect to a single parameter. While this might not be a realistic modelling framework for ordinal outcomes in general, it is a welcome simplification in the absence of prior information on the mechanism of the effect with respect to different levels of disease severity. For an example of a more robust variation of the PO model, see \cite{MurYuaTha18}.

For the remainder of the article, we focus on $\mathbf{p}$ and $\mathrm{OR}$ rather than $\alpha_k$ and $\beta$ as the model parameters since the formulation of the hypotheses and the baseline assumptions are made for these parameter transformations. Specifically, the null and alternative hypotheses are  $H_0: \mathrm{OR}\geq 1$ and $H_A: \mathrm{OR}< 1.$

The trial hypothesis may be defined according to a minimum clinically important effect, e.g., $\mathrm{OR}<0.9$. However, even a small effect was clinically important in the PROTECT study. The methods described in Section~\ref{Sec:methods} apply to generic $H_A$.

For making Bayesian inference, however, prior distributions are specified for the $\alpha_k$ and $\beta$, and the inference is based on the posterior distribution of these parameters given the observed data $\mathbf{y}$. This posterior distribution is analytically intractable and needs to be approximated or estimated by MCMC. We use the R package developed in \cite{bayesCPM} that uses Hamiltonian Monte Carlo implemented in Stan to sample from the posterior distribution of the PO model parameters. The posterior distributions of $\mathbf{p}$ and $\mathrm{OR}$ are obtained respectively as transformations of the posterior samples of $\alpha_k$ and $\beta$.

The posterior probability of effectiveness is then
\begin{equation*}
%	\label{ps}
	\pi = P(H_A\mid \mathbf{y}) = P(\mathrm{OR}<1\mid \mathbf{y}).
\end{equation*}
We focus on the probability of stopping the trial early for efficacy or futility at the interim analysis. Superiority and futility decisions are made if $\pi>U$ or $\pi<\ell$, where $U$ and $\ell$ are prespecified upper and lower probability thresholds, respectively. The  DOCs of interest are therefore the probabilities of these events under the data (sampling) distribution for a variety of model parameter values under the null and alternative hypotheses. For example, the false-positive rate is $P_\mathbf{y}(\pi>U \mid \mathbf{p}_T = \mathbf{p}*, \mathrm{OR}_T \geq 1)$,
where $\mathbf{p}_T$ and $\mathrm{OR}_T$ are the assumed ``true" values of the model parameters.

As explained in Section~\ref{Sec:methods}, the key to estimating probabilities of this form is to estimate the sampling distribution of $\pi$ throughout the parameter space given an initial set of samples drawn from this sampling distribution via trial simulation at a training set. Denote the transformed parameter space for the PO model by $\Theta = \mathcal{P}\times O$, where $\mathcal{P}$ is the four-dimensional simplex of all plausible values of the vector of probabilities $\mathbf{p}$ and $O$ is the interval of $\mathrm{OR}$ values that can be realistically assumed as the effect of the intervention. 

Generating a space-filling design on $\Theta$ is challenging since $\mathcal{P}$ is a simplex defined by the constraints
$\sum_{k = 1}^4 p_k = 1$ and
$ p_{lk}<p_k<p_{uk}$
where the $p_{lk}$s and $p_{uk}$s determine realistic ranges of values for the risk associated with each category. In Section~\ref{sec:des}, we describe a simple design algorithm that generates a space-filling design on $\mathcal{P}$. The design over $\mathcal{P}$ is then combined with a set of equally spaced $\mathrm{OR}$ values to provide training and test sets over the parameter space $\Theta$.

\subsection{Space-filling design over the simplex parameter subspace}
\label{sec:des}

Most space-filling design algorithms, such as those mentioned above, assume a rectangular input space. However, the problem of generating designs on nonrectangular spaces has recently received more attention \citep{LinShaWin10, LekJon2014, MakJos18}. We use the method of \cite{LekJon2014}, which generates and clusters a covering sample of the target space. The cluster centroids are used as the design points. This strategy assures that no two points are too close to each other. Generating a uniform covering sample on a simplex, however, is not straightforward. We propose a sequential Monte Carlo sampling technique that generates an initial sample on the simplex as an efficient and effective alternative to simple Monte Carlo.

Consider the constrained space $\mathcal{P}\subseteq [0,1]^4$. Generating a uniform sample over $\mathcal{P}$ is equivalent to sampling from 
\begin{equation*}\label{eq:target}
	\pi^\top(\mathbf{p})=\frac{\mathcal{U}(\mathbf{p})\mathbf{1}_{\mathcal{P}}(\mathbf{p})}{\int_{\mathcal{P}}\mathcal{U}(\mathbf{p})d\mathbf{p}},
\end{equation*}
where $\mathcal{U}(\cdot)$ is the density function of a uniform distribution with the domain  $[0,1]^4$ and $\mathbf{1}_{\mathcal{P}}(\mathbf{p})$ is an indicator function that is equal to one if $\mathbf{p}\in \mathcal{P}$ and is zero otherwise. 

We use the sequentially constrained Monte Carlo (SCMC) sampling approach proposed in \cite{GolCam16} to sample from $\pi^\top(\mathbf{p})$. Specifically, we define  the deviation of a given point $\mathbf{p}$ from the constraints that define the design region $\mathcal{P}$ as
$$C_{\mathcal{P}}(\mathbf{p}) = \left(\left|\sum_{k = 1}^4p_k - 1\right|, \left(p_k - p_{uk}\right)_{k = 1}^4, \left(p_{lk} - p_k\right)_{k = 1}^4\right).$$
For any point $\mathbf{p}\in\mathcal{P}$, the first component of the deviation vector is zero and the rest are negative. For any point  $\mathbf{p}\notin\mathcal{P}$, $C_{\mathcal{P}}(\mathbf{p}) $ measures the deviation from each of the constraints. 

A probabilistic version of the constraint indicator $\mathbf{1}_{\mathcal{P}}(\mathbf{p})$ is
$\prod \Phi(-\tau C_{\mathcal{P}}(\mathbf{p}))$,
where $\Phi$ is the cumulative distribution function for the standard normal distribution and the parameter $\tau$ controls the slope of the probit function. The number of terms in the  product is equal to the number of constraints defined by $C_{\mathcal{P}}(\mathbf{p})$, which is nine in the present example. We have that
\begin{equation*}\label{eqn:profit}
	\lim_{\tau\rightarrow \infty}\prod \Phi(-\tau C_{\mathcal{P}}(\mathbf{p}))\propto\mathbf{1}_{\mathcal{P}}(\mathbf{p}).
\end{equation*}
From a uniform sample of size $N$ on the unit hypercube $[0,1]^4$, the goal is to filter this sample towards the simplex, $\mathcal{P}$ through a sequence of increasingly constrained densities. 
The SCMC sequence of densities is
\begin{equation*}
	\pi^t(\mathbf{p})\propto \mathcal{U}(\mathbf{p})\prod \Phi(-\tau_t C_{\mathcal{P}}(\mathbf{p})), 
\end{equation*} 
where $ t = 0, \ldots, T$ and $0=\tau_0<\tau_1<\cdots<\tau_T\rightarrow \infty$.

An effective sequence of constraint parameters (the $\tau_t$s) can be achieved adaptively  \citep{JasSteDou11}. At each step, the next value in the sequence is determined such that the effective sample size (ESS) does not fall below a certain threshold, e.g., half of the sample size, $N/2$. This is done by numerically solving for $\tau_t$ in
\begin{equation*}
	\label{eqn:adapt_ESS}
	\text{ESS}=\frac{\left(\sum_{n=1}^N w^{t}_n(\tau_t)\right)^2}{\sum_{n=1}^N\left(w^{t}_n(\tau_t)\right)^2}=\frac{N}{2},
\end{equation*}
where
\begin{equation*}
	\label{eqn:adapt_weight}
	w^{t}_n(\tau_t)=\frac{\prod\Phi(-\tau_t C_{\mathcal{P}}(\mathbf{p}_n^{t-1}))}{\prod\Phi(-\tau_{t-1} C_{\mathcal{P}}(\mathbf{p}_n^{t-1}))}
\end{equation*}
and  $\tau_T = 10^6$. 

\begin{figure}
	%\centerline{\epsfbox{pairs.eps}}
	\includegraphics[width=\textwidth]{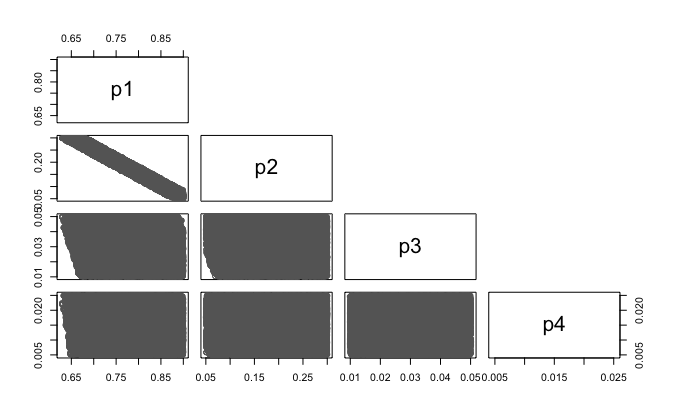}
	\caption{Two-dimensional marginals of the sample generated over the simplex $\mathcal{P}$.}
	\label{fig:p0}
\end{figure}

Figure~\ref{fig:p0} shows the two-dimensional marginal samples generated over $\mathcal{P}$ using the SCMC sampling scheme. The lower and upper limits for $\mathbf{p}$ are arbitrarily specified as $\mathbf{p}_l = (0.5, 0.05, 0.01, 0.005)$ and $\mathbf{p}_u = (0.9, 0.30, 0.05, 0.025)$, respectively. In practice, any information on credible values for the base risk of each disease severity level should be used to specify these thresholds. Note that $\mathbf{p}_l$ and $\mathbf{p}_u$ do not need to belong to $\mathcal{P}$, i.e., the limit vectors do not need to satisfy the constraint that $\sum_{k = 1}^4p_k = 1$.

once a covering sample over the target space is generated, we use the method proposed in \cite{LekJon2014}, which uses K-means clustering and selects the cluster centroids as the design points. This sampling-based approach aims to generate a design in which no two points are too close to each other without relying on a distance measure such as the Euclidean distance, which does not realistically represent distances in a constrained subspace. Various other methods have been developed in the experimental design literature that can be employed here \citep{WelBucSac1996, DraSanDea12, JosGul15, GomClaSer18}. Simplicity of implementation motivates our choice here.

\subsection{Predicted probability of stopping early}
\label{Sec:res}
We consider the PROTECT study design with an interim sample size of $1000$ and estimate the probabilities of stopping early at the interim analysis for efficacy and futility for a range of baseline risks $\mathbf{p}$ and $\mathrm{OR}$. We can similarly estimate power (the probability of efficacy at the interim or final analysis) or any other  DOCs for more complex designs with multiple interim analyses and adaptive features. Design complexities will only need to be implemented in initial simulations and do not have any implications on the prediction methods thereafter. 

The goal is to predict the probabilities that 
\begin{equation}
	\label{eq:sup}
P(\mathrm{OR}<1\mid \mathbf{y})>U
\end{equation}
and
\begin{equation}
	\label{eq:fut}
P(\mathrm{OR}<1\mid \mathbf{y})<\ell
\end{equation}
over the input space $\Theta = \mathcal{P}\times O$ where 
\begin{align*}
\mathcal{P} = \{ &\mathbf{p}: \sum_{k = 1}^4 p_k = 1, \hskip 2pt 0.5 <p_1<0.9, \hskip 2pt 0.05<p_2<0.3, \hskip 2pt0.01<p_3<0.05,\\  &0.005<p_4<0.025\}
\end{align*}

For the training set, a space-filling design of size 20 is generated over $\mathcal{P}$ using the methods described in Section~\ref{sec:des}. The Cartesian product of this set and a grid of size four over $O =  (0.7, 1)$ is used as the final design of size 80 over $\Theta$. Trial data are then simulated from the PO model described in Section~\ref{Sec:PO} for each of the $(\mathbf{p}, \mathrm{OR})$ pairs in the training set. For every simulated trial, the probability of superiority of the treatment (given in Equation (\ref{ps})) is obtained to estimate the sampling distribution of $\pi$. A beta density function is fit to the Monte Carlo samples of $\pi$ at every $\theta = (\mathbf{p}, \mathrm{OR})$ to obtain estimates of $a(\theta)$ and $b(\theta)$. GPs are then fit to these simulated values.

\begin{figure}
	%\centerline{\epsfbox{power_pred_obs.eps}}
	\centering
	\begin{subfigure}[b]{0.8\textwidth}
	\centering
	\includegraphics[width=\textwidth]{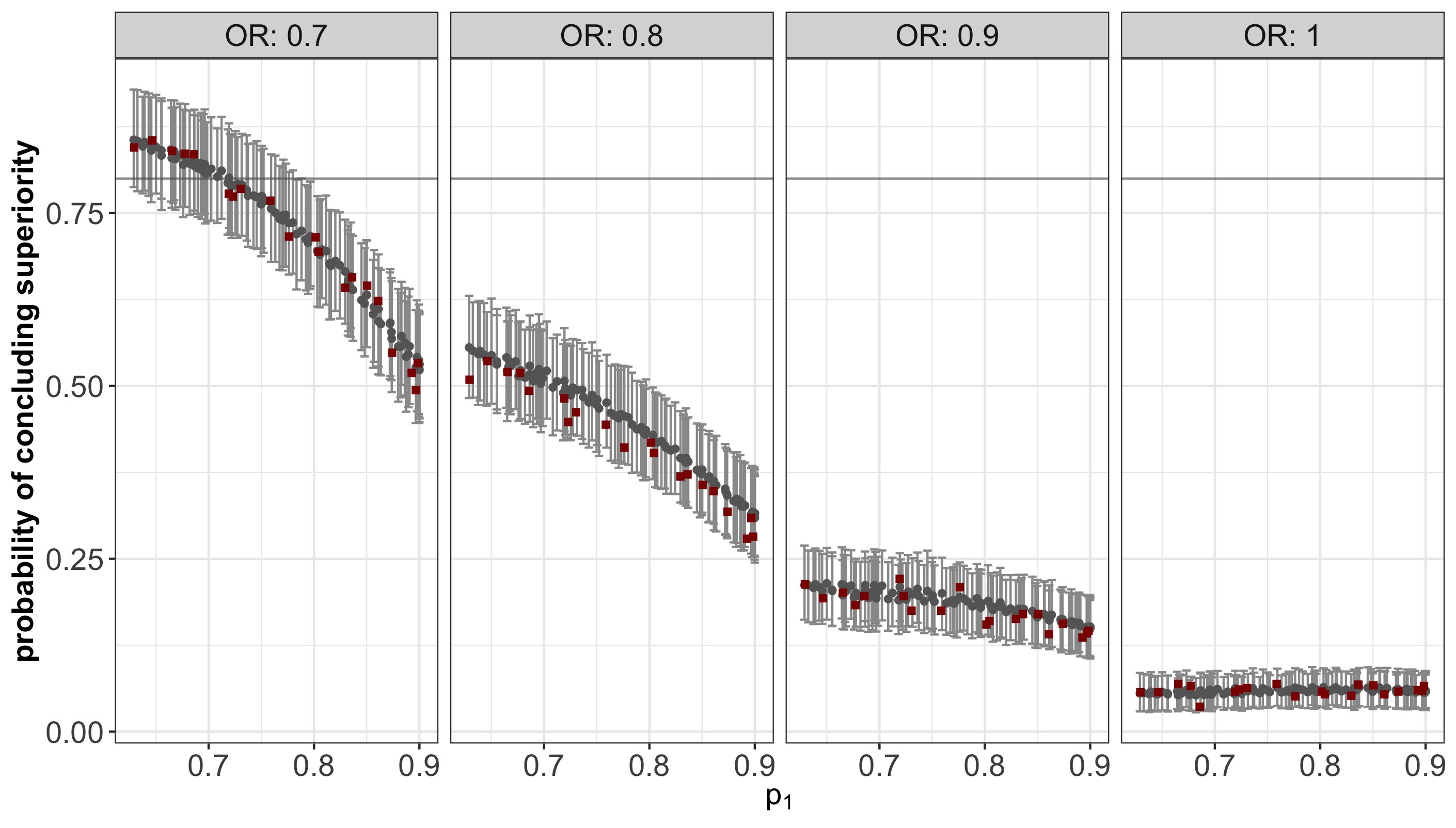}
	\caption{}
	\label{fig:01}
\end{subfigure}
\begin{subfigure}[b]{0.8\textwidth}
	\centering
	\includegraphics[width=\textwidth]{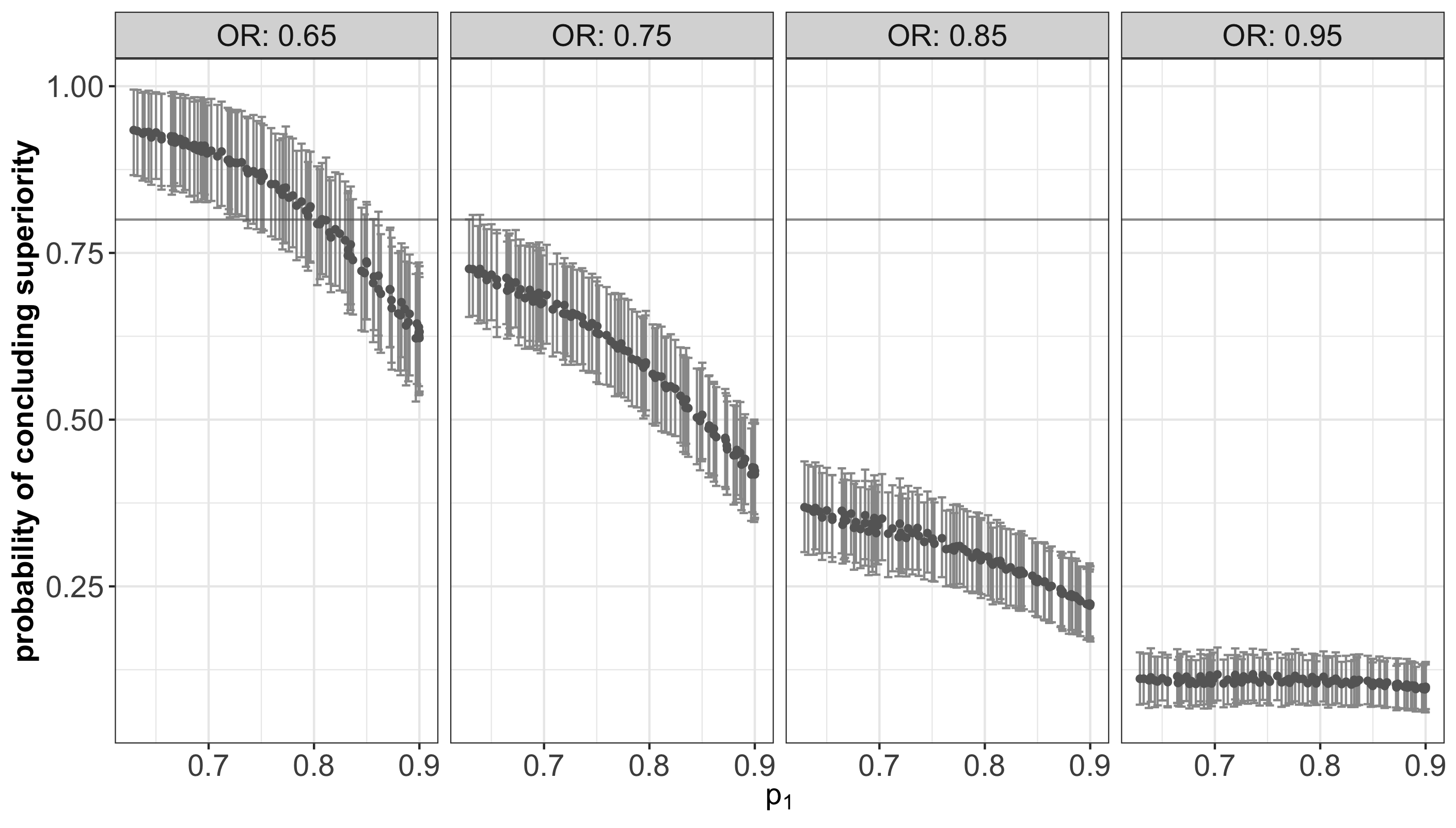}
	\caption{}
	\label{fig:02}
\end{subfigure}
	\caption{Point estimates for the interim probability of concluding superiority (grey round dots) with 95\% credible intervals.  Panel (a) shows a slice of the subspace where the training set (denoted by square dots) is located. Panel (b) shows a slice of the subspace ``in between" the training points.}
	\label{fig:interimsup}
\end{figure}

The sampling distribution of $\pi$ is then predicted over a test set of size 800 generated over $\Theta$ in the same fashion as the training set. Figures~\ref{fig:01} and \ref{fig:02} show the predicted probability of stopping early for efficacy (using a decision threshold of $U = 0.95$) derived from the distribution of $\pi$ together with 95\% credible intervals over a two-dimensional subspace of $\Theta$, i.e., $p_1\times \mathrm{OR}$. The upper panel shows a slice of the subspace where the training set (denoted by square dots) is located while the lower panel shows a slice of the subspace ``in between" the training points. The 95\% credible intervals represent two layers of uncertainty, i.e., the observation (Monte Carlo) error of the initial  DOCs evaluations and the uncertainty associated with the estimation/prediction step in obtaining the sampling distribution of the test statistic and the  DOCs as its quantile.

As mentioned earlier, a strength of the proposed approach is that it obtains the sampling distribution of the test statistic rather than focusing on a single operating characteristic corresponding to a fixed decision criterion. This means that once the model is trained, one can readily explore a variety of decision thresholds $U$ in (\ref{eq:sup}) without any additional simulation runs. 

\begin{figure}
	\centering
	\begin{subfigure}[b]{0.8\textwidth}
		\centering
		\includegraphics[width=\textwidth]{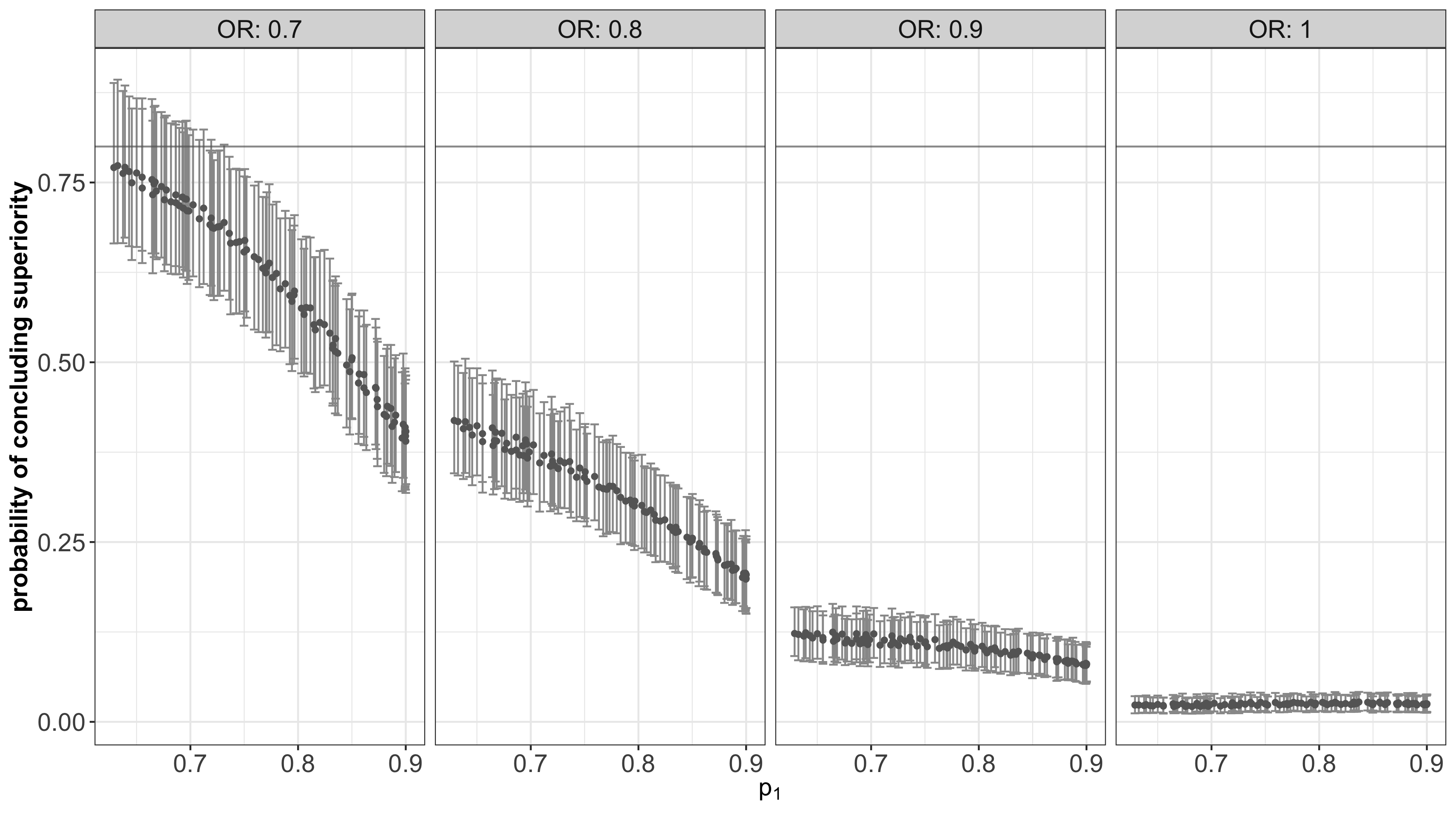}
		\caption{}
		\label{fig:98}
	\end{subfigure}
	\begin{subfigure}[b]{0.8\textwidth}
		\centering
		\includegraphics[width=\textwidth]{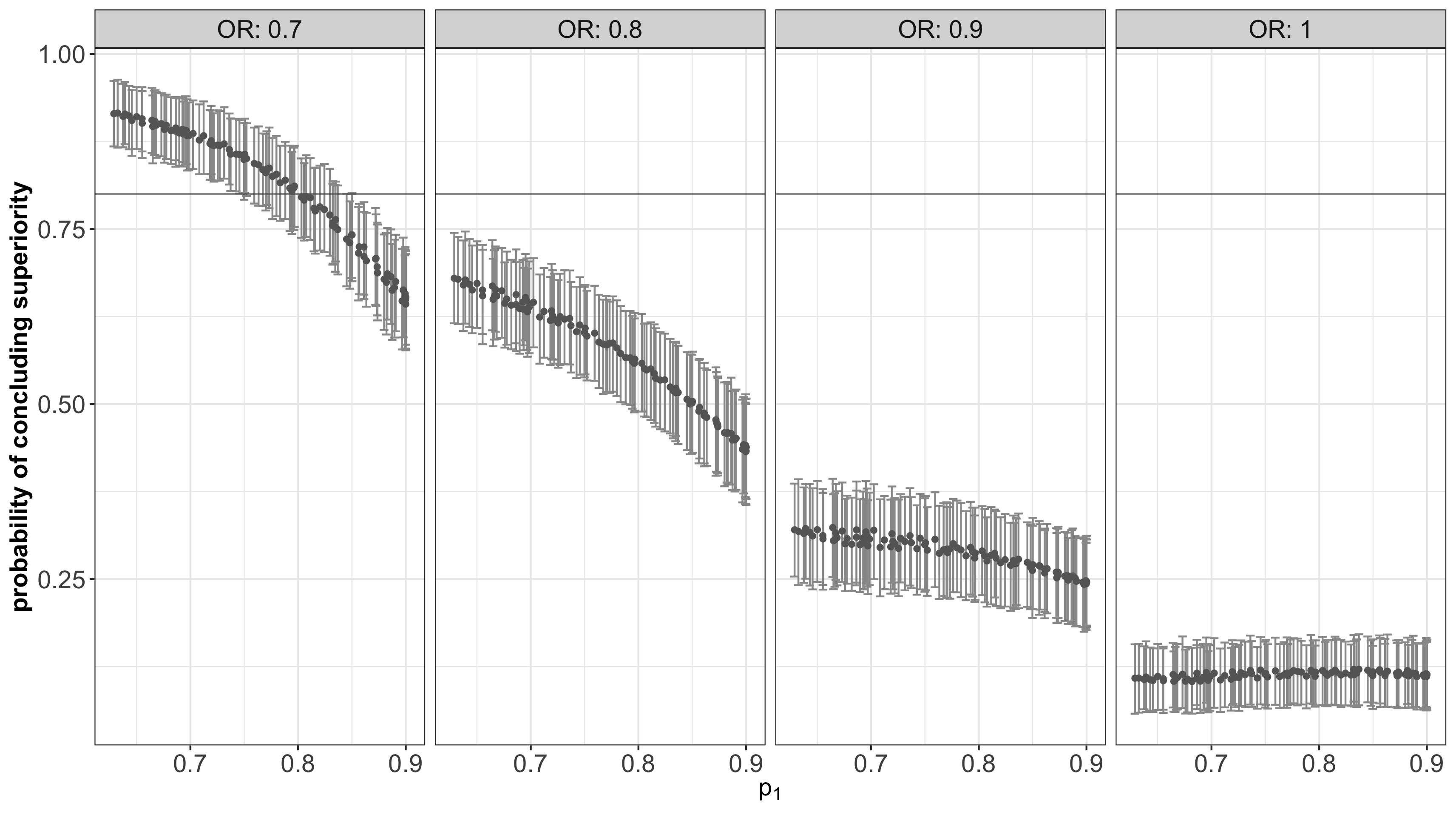}
		\caption{}
		\label{fig:90}
	\end{subfigure}

	\caption{Point estimates and 95\% credible intervals for the probability of concluding superiority with the decision thresholds (a) 0.98 and (b) 0.9.}
	\label{fig:9890}
\end{figure}

To showcase this feature, Figure~\ref{fig:9890} shows the results in Figure~\ref{fig:01}---excluding the training points---for the decision thresholds $U=0.98$ and $U=0.9$. These values correspond to more- and less-conservative decision rules relative to $U=0.95$, respectively. Using the more-conservative decision criterion ($U = 0.98$) results in a smaller probability of stopping early for superiority for $\mathrm{OR}<1$, but controls the probability of a false positive result at the interim analysis (Figure~\ref{fig:98}, where $\mathrm{OR} =1$). The more-permissive decision threshold ($U = 0.9$), however, increases the chance of stopping the trial early due to efficacy but increases the chance of a false positive result to 12.5\% (Figure~\ref{fig:90}, where $\mathrm{OR} =1$).

\begin{figure}
	\centering
	\begin{subfigure}[b]{0.8\textwidth}
		\centering
		\includegraphics[width=\textwidth]{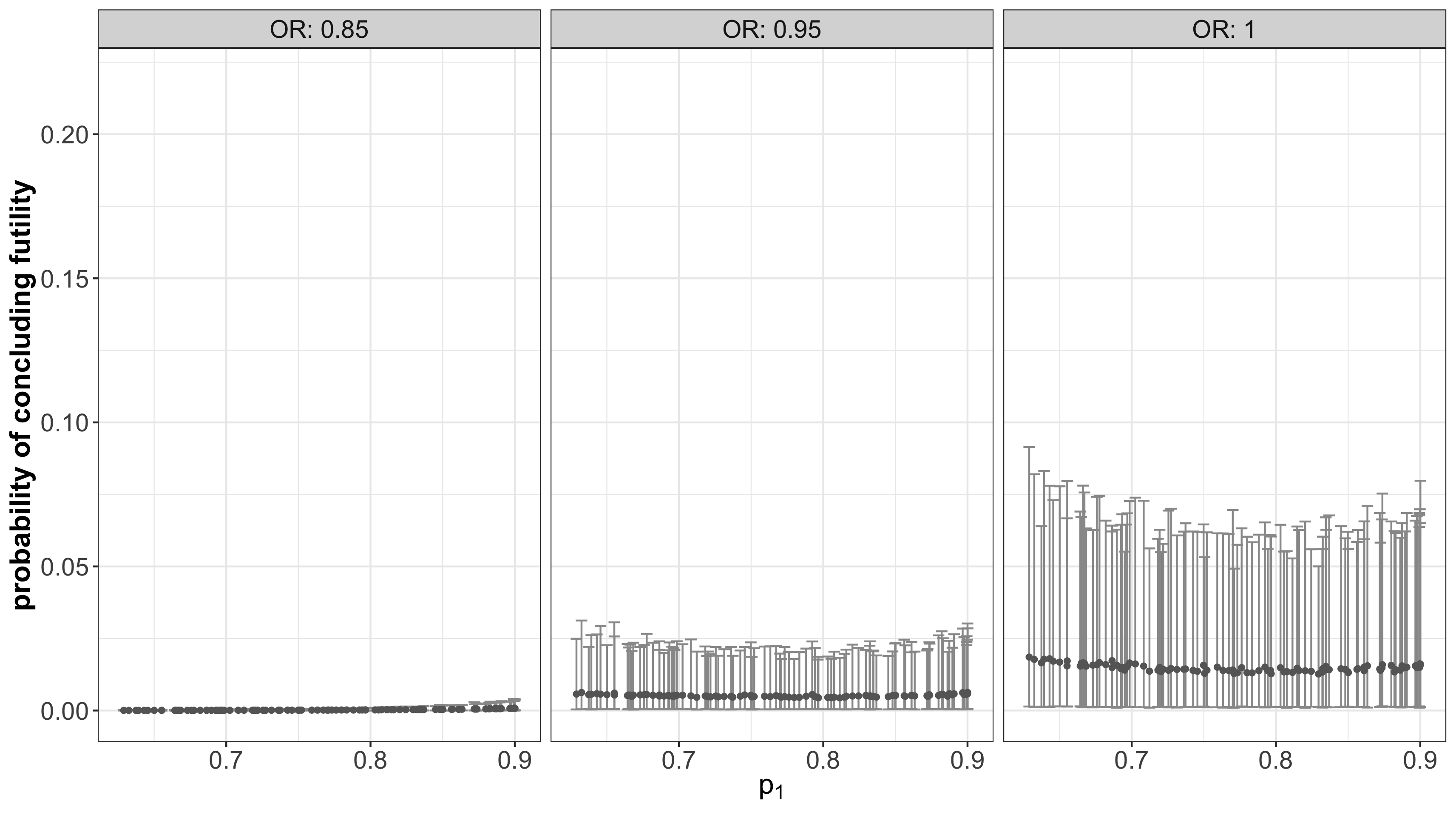}
		\caption{}
		\label{fig:fut01}
	\end{subfigure}
	\begin{subfigure}[b]{0.8\textwidth}
		\centering
		\includegraphics[width=\textwidth]{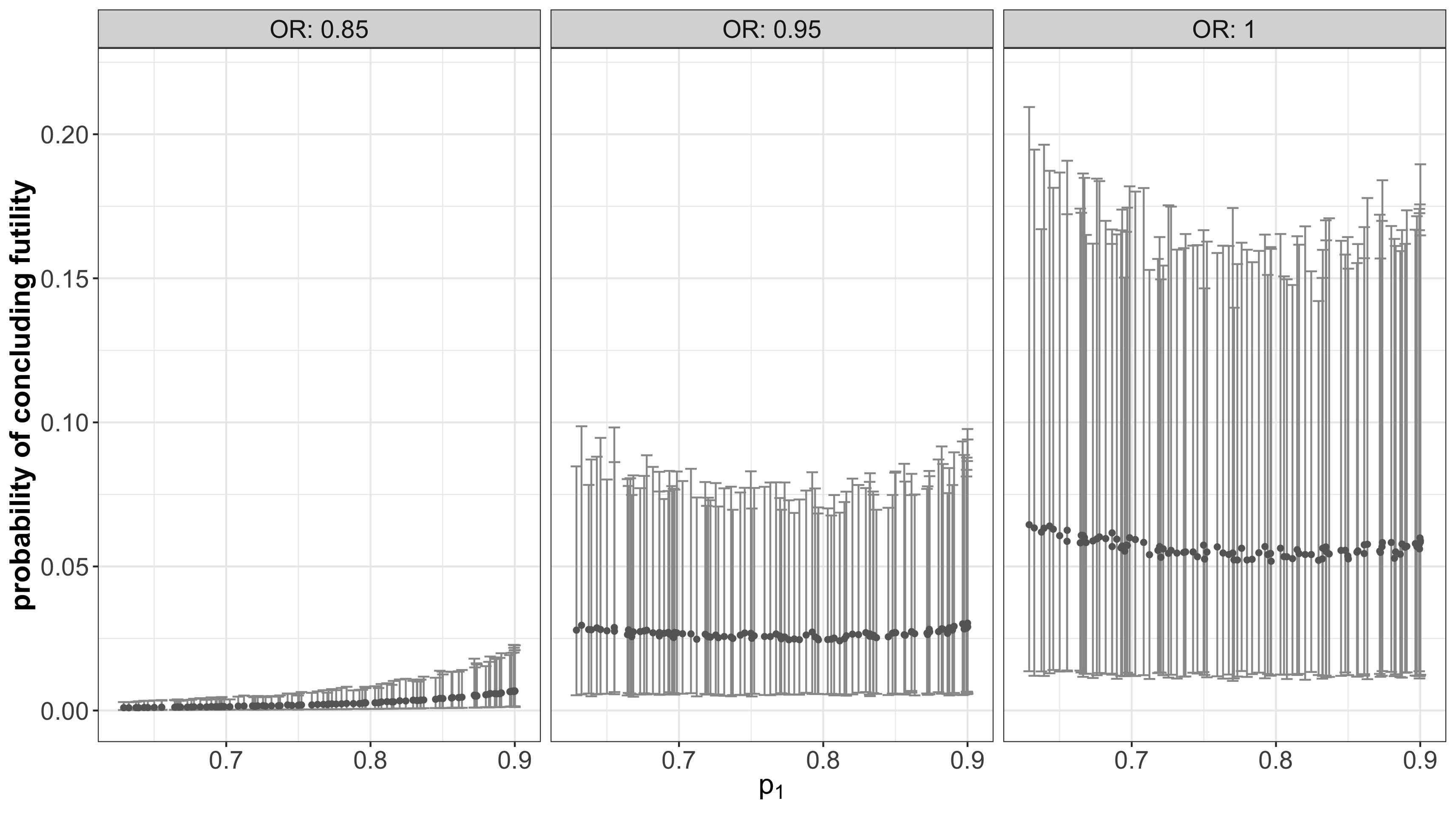}
		\caption{}
		\label{fig:fut05}
	\end{subfigure}
	
	\caption{Point estimates and 95\% credible intervals for the probability of concluding futility with the decision thresholds (a) 0.01 and (b) 0.05.}
	\label{fig:fut}
\end{figure}

Likewise, we may explore other operating characteristics such as the probability of concluding futility, according to (\ref{eq:fut}) for various decision thresholds $\ell$. Figure~\ref{fig:fut} shows the probabilities of concluding futility at the interim analysis for a restricted range of parameter values. This probability is negligible for larger effect sizes with the specified design settings.

The results in Figures~\ref{fig:interimsup}--\ref{fig:fut} help to select a set of decision criteria for a given set of model parameters corresponding to baseline risks and effect assumptions. For example, if a base risk vector of $\mathbf{p}=(0.75, 0.22, 0.01, 0.02)$ and $\mathrm{OR} = 0.7$ is assumed, then to allow a 65\% (CI: 56\%--75\%) chance of stopping the trial early for superiority while controlling the interim false positive rate at 2.3\% (CI:1.3\%--3.5\%) requires a superiority decision threshold of $U = 0.98$.  For the same set of assumptions for $\mathbf{p}$ but assuming no effect ($\mathrm{OR} =1$) a futility decision threshold of $\ell = 0.05$ allows for about a 5\% (CI: 1.1\%--16\%) chance of stopping the trial early for futility at the interim analysis.

\subsection{Computational savings}
We next present the computational savings in the application of the proposed approach in the previous subsection. All operations were performed in serial. Clearly, further savings can be attained through parallel computing. We argue, however, that significant computational resources are required to achieve the same level of savings through only parallel computing. 

On a 1.4 GHz Quad-Core Intel Core i5 processor, one trial simulation with one chain and 1000 Hamiltonian Monte Carlo iterations took slightly over two seconds. Therefore, 1000 trial simulations for one set of model parameters would take about 30 min if run in serial. For 80 points in the training set, this takes about 40 h and for the 800 points in the test set, 400 h. The proposed approach relies on the 80-training-point simulation and therefore includes the 40 h of computation time. However, it takes about 20 s to fit each GP model (2 chains and 2000 iterations each), i.e., less than a minute in total. Sampling from the posterior predictive distribution for the test set takes about a minute. Therefore, the computation time for making predictions on the test set is reduced from 400 h to two min. To reduce the computation time via parallel computing alone to a comparable level would require significant computational resources (thousands of processing units). Of course, parallel computing together with the proposed methods can result in further reductions in computation times.

\subsection{Assessment of predictive performance via cross-validation}
In this section we provide an assessment of the accuracy and precision of the proposed approach in estimating DOCs relative to trial simulation. We use the training set $\mathbf{\theta}_t$ (of size $n_t = 80$) used in the previous section to calculate the root mean squared error (RMSE) using leave-one-out cross-validation: 
\begin{equation*}
	\text{RMSE} = \sqrt{\frac{1}{n_tK}\sum_{i = 1}^{n_t}\sum_{k = 1}^K (\phi_k(\theta_i) - \phi_t(\theta_i))^2}.
\end{equation*}
Here, $\phi_k(\theta_i) = P(\pi>0.95 \mid a_k(\theta_i), b_k(\theta_i))$ is the estimate of the interim probability of superiority obtained as an upper tail probability of a beta distribution with the parameters given by the $k^\text{th}$ posterior samples $a_k(\theta_i)$ and  $b_k(\theta_i)$ drawn from GP posteriors trained over the $n_t - 1$ points in $\mathbf{\theta}_t$ with $\theta_i$ excluded, and $\phi_t(\theta_i)$ is the ``true" interim power obtained via trial simulation at $\theta_i$. 

\begin{figure}
	\centering
	\begin{subfigure}[b]{0.8\textwidth}
		\centering
		\includegraphics[width=\textwidth]{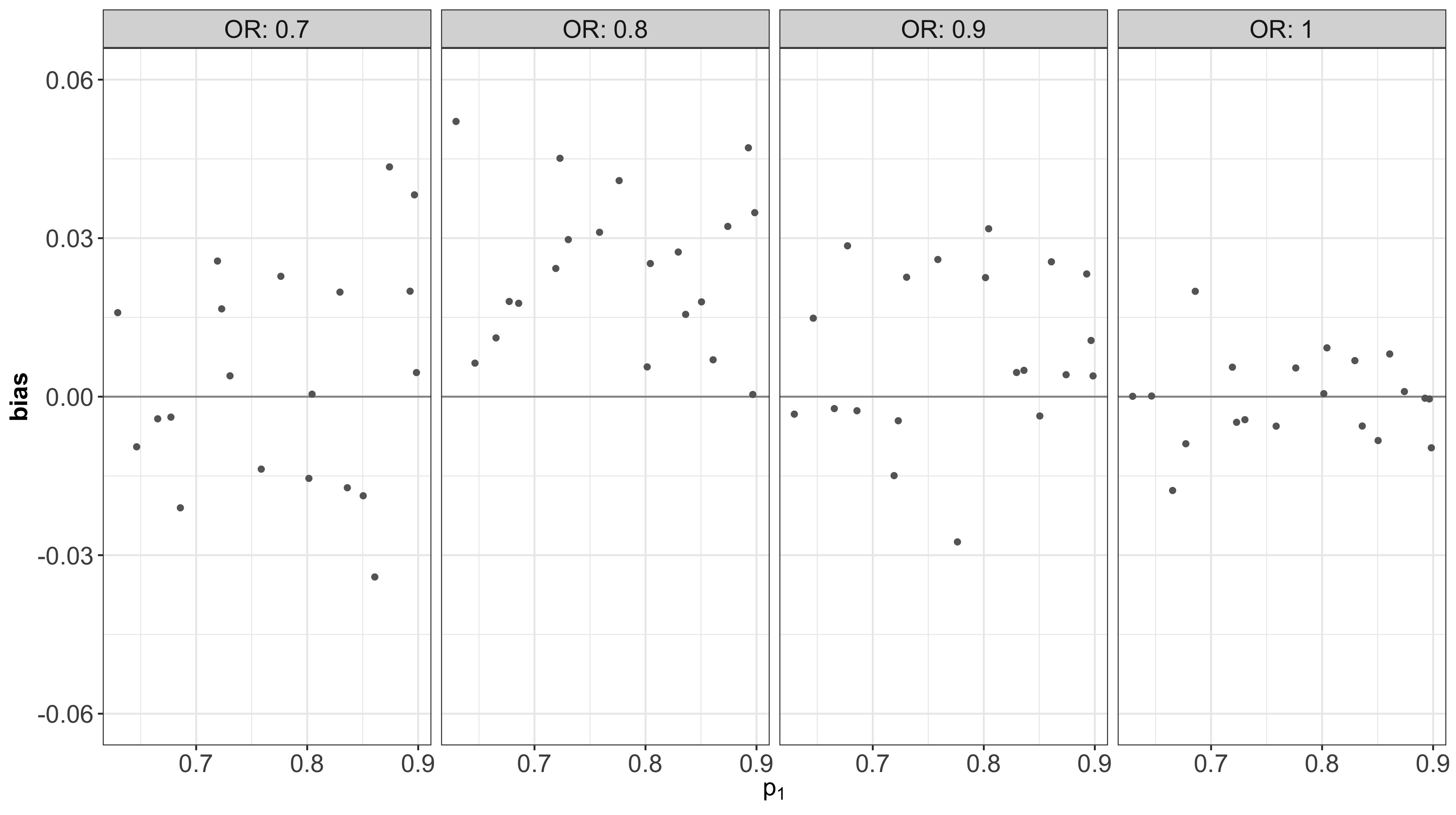}
		\caption{}
		\label{fig:bias_CV}
	\end{subfigure}
	\begin{subfigure}[b]{0.8\textwidth}
		\centering
		\includegraphics[width=\textwidth]{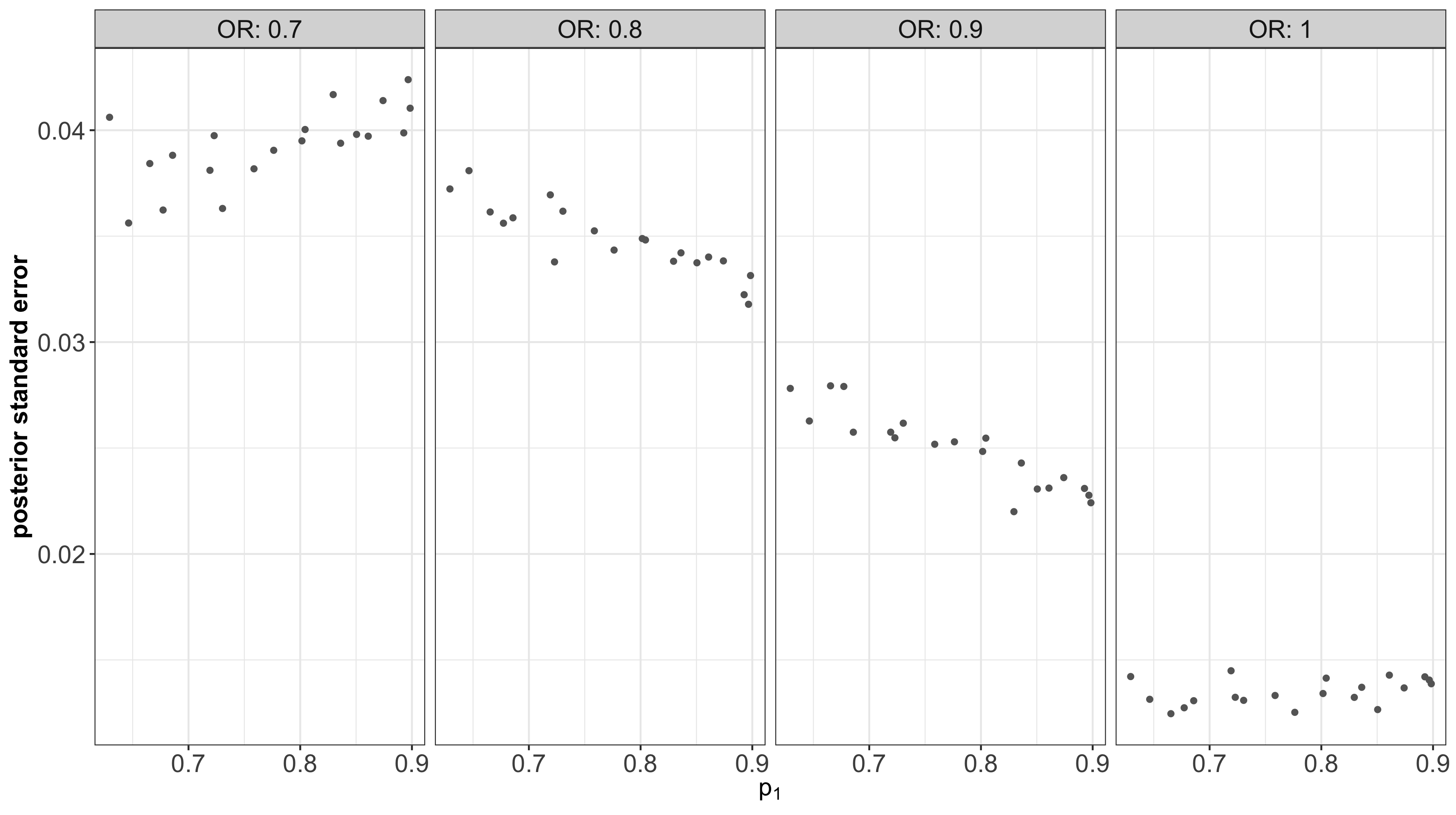}
		\caption{}
		\label{fig:pse_CV}
	\end{subfigure}
	
	\caption{Cross-validated (a) bias and (b) posterior standard error in estimates of the probability of stopping early for 80 points across the parameter space.}
	\label{fig:CV}
\end{figure}

The RMSE, as defined above, is averaged both over the posterior distribution and the parameter space and is evaluated at 0.036. The estimation error---including both bias and variance---varies across the parameter space. Therefore, assessing pointwise accuracy and precision is important. Figure~\ref{fig:CV} shows the cross-validated bias and posterior standard error for the 80 points in the training set.

While estimation bias is small in magnitude overall (Figure~\ref{fig:bias_CV}) there appears to be a systematically positive bias in the $\mathrm{OR}\approx0.8$ region of the parameter space. This is the region where the sampling distribution changes and as is discussed in Section~\ref{Sec:dis}, the beta distribution with stationary GP prior distributions is not able to adequately capture this change. The same phenomenon contributes to the large posterior variance in these regions (Figure~\ref{fig:pse_CV}). However, the estimation error remains small and the 95\% credible intervals provide 100\% coverage as indicated in Figure S.1 of the supplementary material.

For a more-extensive assessment of the proposed approach that takes into account the sensitivity of the predictions to the design of the training set, see the simulation study provided in the supplementary material. The simulation study is designed within the simple framework of a binary outcome with a beta--binomial model to be able to take advantage of the conjugate modelling framework and the consequent analytic posterior distribution. Because of the analytic posterior distribution, the  DOCs can be estimated for a large number of points across the parameter space and for multiple training sets arising from the random training set design. This allows a thorough exploration of the accuracy and precision of  DOCs estimates. The results of the simulation study are consistent with the conclusions drawn from the cross-validated performance measures above.

\section{Discussion}
\label{Sec:dis}

In this article we propose a set of methods for estimating the sampling distribution of a Bayesian probability statement---used for decision making in Bayesian adaptive trials---over a model parameter space. Our goal is to estimate a variety of DOCs and to assess their sensitivity to trial assumptions and design configurations in an efficient manner. We take advantage of the spatial correlation throughout the model parameter space to interpolate the parameters of the sampling distribution. We model the parameters of the sampling distribution  as independent GPs trained over a set of estimates obtained by simulating the trial design over an initial set of model parameter values.

The main advantage of the proposed approach is that it enables exploration of a variety of operating characteristics as well as adequate uncertainty quantification. The methods presented in this article add efficiency to the overall process of collaborative trial design (possibly involving several iterations resulting from proposed changes and extensive explorations) in two ways: first, by reducing the number of simulations required at each iteration of the process following a major change to the design and, second, by eliminating the need for additional simulations if the changes are only through model parameters or decision thresholds. This feature is showcased in Section~\ref{Sec:PO}, where the probabilities of stopping early for superiority or futility are estimated for a variety of decision thresholds with no additional simulation runs.

We apply the proposed methods to a hypothetical clinical trial design where an ordinal scale disease progression endpoint is used with the PO model. The vector of base risks associated with the levels of the ordinal outcome is defined as a simplex contained within the unit hypercube. This results in unique challenges in the design of the initial set of simulations used for training the GPs and in the exploration of  DOCs estimates across the parameter space. We describe a specialized design algorithm for this problem.

This article sets the foundation for the development of methods that facilitate the adoption of Bayesian measures for decision-making in clinical trials. However,  the proposed approach goes beyond trials that employ Bayesian decision rules and may be used generally where the sampling distribution of a test statistic is not available in an analytic form. For example, \cite{BarVilGey21} recently proposed a novel statistical test in clinical trials with response adaptive randomization whose sampling distribution of the test statistic needs to be estimated by simulation: this approach can benefit from the methods proposed in this article. The proposed methodology is also applicable where the sampling distribution of the test statistic is only available asymptotically (which is the case for most frequentist tests) or only for large samples at interim analyses \citep{HadHirZha21}.

We acknowledge that this work is a starting point that has limitations. There remains much room for further developments. We lay out some directions for future work below.

The parameters of the beta distribution must be positive. Therefore, specifying a GP prior that assigns nonzero probabilities to negative values is problematic. We address this issue using rejection sampling  from the posterior predictive distribution. For the application in this article the rejection rates remain zero or very small throughout the parameter space (see Section C in the supplementary materials).  

In cases where rejection sampling is inefficient, however, constraints need to be incorporated into the model. A variety of methods have been proposed for fitting GP models to observations from constrained functions \citep{RiiVeh10, LinDun14, GolBinChi15, WanBer16}. As a common approach, the warped GP \citep{SneRasGha04} maps observations onto the real line via a monotonic function. All existing methods create a non-Gaussian process prior over the original function for which analytic predictions conditional on GP parameters cannot be obtained. Given the practical efficiency of a simple rejection-sampling scheme for the application in this paper and the simplicity of unconstrained GP models, we did not find the additional complexities in incorporating positivity constraints to be justifiable. Doing so may be necessary in different modelling settings, however. 

Another challenging feature of the problem is the quickly changing form of the sampling distribution of the test statistic in certain parts of the parameter space. The GP is not the most appropriate model for the parameters of this distribution as this assumes stationarity throughout the input space. This can lead to increased bias in certain parts of the model parameter space. A variety of methods have been proposed to model nonstationary response surface \citep{SchOha03, GraLee08, GraApl13, HeiManRou16}. In most cases, however, the solution comes at the cost an analytic form for the predictions or uncertainty estimates and an increased computational burden. In the examples explored in the present article, the resulting bias is small enough, so we consider the fit of a conventional GP to be satisfactory. 

%Another aspect of the proposed methodology that may be improved is incorporating the constraints into the GP fit rather than correcting for them as a post processing step in sampling from the posterior predictive distribution. Note that both the shape and scale parameters of the beta distribution used as the sampling distribution of the test statistic are bounded to be positive and are monotonic with respect to at least one of the model parameters (Figures~\ref{fig:alpha} and \ref{fig:beta}). As mentioned earlier, various techniques exist for incorporating constraints and boundary information into GP models \citep{Tan2018, SolKok19, DinMakWu19}. However, similar to the argument made above, when employing these methods one must assure that the additional computational complexity and potential loss of generality is justified by the gain in estimation accuracy and precision.
%
%\begin{figure}[t]
%	
%	\begin{subfigure}[b]{0.95\textwidth}
%		\centering
%		\includegraphics[width=\textwidth]{alpha.png}
%		\caption{}
%		\label{fig:alpha}
%	\end{subfigure}
%	\begin{subfigure}[b]{0.95\textwidth}
%		\centering
%		\includegraphics[width=\textwidth]{beta.png}
%		\caption{}
%		\label{fig:beta}
%	\end{subfigure}
%	
%	\caption{Point estimates and 95\% credible intervals for the (a) shape and (b) scale parameters of the sampling distribution of the posterior probability of superiority }
%	\label{fig:ab}
%\end{figure}

Finally, the parametric form assumed for the Bayesian test statistic, e.g., the beta distribution alone might not capture the true sampling distribution throughout the model parameter space. A nonparametric modelling approach is therefore a potential future direction for this work. 
%Specifically, the spatial correlation may be imposed through the dependent Dirichlet processes framework \citep{MacEachern1999, MacEachern2000, QuiMueJar20}.

\section*{Code}
The code for the simulation study as well as the implementation of the methods for the ordinal scale endpoint and the PO model used in this article are provided in a public repository on GitHub, at \href{https://github.com/sgolchi/ DOCsest}{https://github.com/sgolchi/ DOCsest}.

\section*{Acknowledgements}
The author gratefully acknowledges the  two principal investigators of the PROTECT study, Dr. Francine M. Ducharme and Dr. Cecile Tremblay, the review committee and Dr. Alexandra Schmidt for their invaluable comments which resulted in significant improvements to this article. This work is supported by a Discovery Grant from the Natural Sciences and Engineering Research Council of Canada (NSERC).

\bibliographystyle{apalike}
\bibliography{ref01}

\end{document}